\documentclass[12pt,preprint]{aastex}
\usepackage{wrapfig}
\usepackage{natbib}

\newcommand{\kms}{\ifmmode {\rm km~s}^{-1} \else km~s$^{-1}$\fi}
\newcommand{\ergs}{\ifmmode {\rm erg~ s}^{-1} \else erg~s$^{-1}$\fi}
\newcommand{\ergscm}{\ifmmode {\rm erg~s}^{-1} \else erg~s$^{-1}$ cm$^{-2}$\fi}
\newcommand{\Msun}{\ifmmode {\rm M}_{\odot} \else M$_{\odot}$\fi }
\newcommand{\Lsun}{\ifmmode {\rm L}_{\odot} \else L$_{\odot}$\fi}
\newcommand{\qo}{\ifmmode q_{\rm o} \else $q_{\rm o}$\fi}
\newcommand{\Ho}{\ifmmode H_{\rm o} \else $H_{\rm o}$\fi}
\newcommand{\ho}{\ifmmode h_{\rm o} \else $h_{\rm o}$\fi}

\newcommand{\vFWHM}{\ifmmode v_{\mbox{\tiny FWHM}} \else
                    $v_{\mbox{\tiny FWHM}}$\fi}
\newcommand{\CCF}{\ifmmode F_{\it CCF} \else $F_{\it CCF}$\fi}
\newcommand{\ACF}{\ifmmode F_{\it ACF} \else $F_{\it ACF}$\fi}
\newcommand{\Halpha}{\ifmmode {\rm H}\alpha \else H$\alpha$\fi}
\newcommand{\Hbeta}{\ifmmode {\rm H}\beta \else H$\beta$\fi}
\newcommand{\Hgamma}{\ifmmode {\rm H}\gamma \else H$\gamma$\fi}
\newcommand{\Hdelta}{\ifmmode {\rm H}\delta \else H$\delta$\fi}
\newcommand{\Lya}{\ifmmode {\rm Ly}\alpha \else Ly$\alpha$\fi}
\newcommand{\Lyb}{\ifmmode {\rm Ly}\beta \else Ly$\beta$\fi}
\newcommand{\HeI}{\ifmmode {\rm He}\,{\sc i}\,\lambda5876 \else 
	          He\,{\sc i}\,$\lambda5876$\fi}
\newcommand{\HeII}{\ifmmode {\rm He}\,{\sc ii}\,\lambda4686 \else 
	           He\,{\sc ii}\,$\lambda4686$\fi}

\newcommand{\heii}{He\,{\sc ii}}

\newcommand{\ciii}{\ifmmode {\rm C}\,{\sc iii} \else C\,{\sc iii}\fi}
\newcommand{\civ}{C\,{\sc iv}}

\newcommand{\oiii}{O\,{\sc iii}}

\newcommand{\mbh}{$M_{\rm BH}$\ }
\newcommand{\sigstar}{$\sigma_{*}$\ }

\shorttitle{A High-Ionization Reverberation Lag Measurement for Mrk 335}
\shortauthors{Grier et al.}

\begin{document}

\title{A Reverberation Lag for the High-Ionization Component of the Broad Line Region in the Narrow-Line Seyfert 1 Mrk 335}

\author{C.~J.~Grier\altaffilmark{1},
B.~M.~Peterson\altaffilmark{1,2},
R.~W.~Pogge\altaffilmark{1,2},
K.~D.~Denney\altaffilmark{3},
M.~C.~Bentz\altaffilmark{4},
Paul~Martini\altaffilmark{1,2},
S.~G.~Sergeev\altaffilmark{5},
S.~Kaspi\altaffilmark{6,7},
Y.~Zu\altaffilmark{1},
C.~S.~Kochanek\altaffilmark{1,2},
B.~J.~Shappee\altaffilmark{1},
K.~Z.~Stanek\altaffilmark{1,2},
C.~Araya~Salvo\altaffilmark{1},
T.~G.~Beatty\altaffilmark{1},
J.~C.~Bird\altaffilmark{1},
D.~J.~Bord\altaffilmark{8},
G.~A.~Borman\altaffilmark{5,9},
X.~Che\altaffilmark{10},
C.~Chen\altaffilmark{11},
S.~A.~Cohen\altaffilmark{11},
M.~Dietrich\altaffilmark{1},
V.~T.~Doroshenko\altaffilmark{5,9,12},
Yu.~S.~Efimov\altaffilmark{5,13},
N.~Free\altaffilmark{14},
I.~Ginsburg\altaffilmark{11},
C.~B.~Henderson\altaffilmark{1},
Keith~Horne\altaffilmark{15},
A.~L.~King\altaffilmark{10},
K.~Mogren\altaffilmark{1},
M.~Molina\altaffilmark{1},
A.~M.~Mosquera\altaffilmark{1},
S.~V.~Nazarov\altaffilmark{5,9},
D.~N.~Okhmat\altaffilmark{5,9},
O.~Pejcha\altaffilmark{1},
S.~Rafter\altaffilmark{7},
J.~C.~Shields\altaffilmark{14},
J.~Skowron\altaffilmark{1},
D.~M.~Szczygiel\altaffilmark{1},
M.~Valluri\altaffilmark{10},
and J.~L.~van~Saders\altaffilmark{1}
}

\altaffiltext{1}{Department of Astronomy, The Ohio State University,
140 W 18th Ave, Columbus, OH 43210} 
\altaffiltext{2}{Center for
Cosmology \& AstroParticle Physics, The Ohio State University, 191
West Woodruff Ave, Columbus, OH 4321, USA} 
\altaffiltext{3}{Dark
Cosmology Centre, Niels Bohr Institute, University of Copenhagen,
Juliane Maries Vej 30, DK-2100 Copenhagen, Denmark}
\altaffiltext{4}{Department of Physics and Astronomy, Georgia State
University, Astronomy Offices, One Park Place South SE, Suite 700,
Atlanta, GA 30303, USA} 
\altaffiltext{5}{Crimean Astrophysical
Observatory, P/O Nauchny Crimea 98409, Ukraine}
\altaffiltext{6}{School of Physics and Astronomy, Raymond and Beverly
Sackler Faculty of Exact Sciences, Tel Aviv University, Tel Aviv
69978, Israel} 
\altaffiltext{7}{Physics Department, Technion, Haifa
32000, Israel} 
\altaffiltext{8}{Department of Natural Sciences, The
University of Michigan - Dearborn, 4901 Evergreen Rd, Dearborn, MI
48128} 
\altaffiltext{9}{Isaac Newton Institute of Chile, Crimean
Branch, Ukraine} 
\altaffiltext{10}{Department of Astronomy, University
of Michigan, 500 Church Street, Ann Arbor, MI 41809}
\altaffiltext{11}{Department of Physics and Astronomy, Dartmouth
College, 6127 Wilder Laboratory, Hanover, NH 03755}
\altaffiltext{12}{Crimean Laboratory of the Sternberg Astronomical
Institute, University of Moscow, Russia; P/O Nauchny, 98409 Crimea,
Ukraine} 
\altaffiltext{13}{Deceased, 2011 Oct 21}
\altaffiltext{14}{Department of Physics \& Astronomy, Ohio
University, Athens, OH 45701} 
\altaffiltext{15}{SUPA Physics and Astronomy, University of St~Andrews, 
Fife, KY16 9SS Scotland, UK}

\begin{abstract}
We present the first results from a detailed analysis of photometric
and spectrophotometric data on the narrow-line Seyfert 1 galaxy
Mrk~335, collected over a 120-day span in the fall of 2010. From these
data we measure the lag in the \HeII \ broad emission line relative to
the optical continuum to be 2.7 $\pm$ 0.6 days and the lag in the
\Hbeta$\lambda4861$ broad emission line to be 13.9 $\pm$ 0.9
days. Combined with the line width, the \heii \ lag yields a black
hole mass, \mbh = ($2.6 \pm 0.8 $)$ \times 10^{7} M_{\odot}$. This
measurement is consistent with measurements made using the
\Hbeta$\lambda4861$ line, suggesting that the \heii \ emission
originates in the same structure as \Hbeta, but at a much smaller
radius. This constitutes the first robust lag measurement for a
high-ionization line in a narrow-line Seyfert 1 galaxy and supports a
scenario in which the \heii \ emission originates from gas in virial
motion rather than outflow.
\end{abstract}


\section{INTRODUCTION}
Narrow-line Seyfert 1 galaxies (NLS1s) are a subset of active galactic
nuclei (AGNs) that show narrower broad emission-line components than
typical Type 1 AGNs, as well as a number of other distinguishing
properties (\citealt{Osterbrock85}, \citealt{Goodrich89},
\citealt{Boller96}). Explanations for their unique characteristics
include the possibility that they are either low-inclination or
high-Eddington rate accreters (or both -- see
\citealt{Boroson11}). Substantial blue enhancements in high ionization
lines such as \civ$\lambda$1549 and \heii$\lambda$1640 are apparently
typical of, although not restricted to, NLS1 galaxies
(e.g., \citealt{Richards02}, \citealt{Sulentic00}). This may be
evidence for material in a disk wind (\citealt{Richards11},
\citealt{Leighly04}). If the blue enhancement is due to a wind, the
use of high ionization emission lines to measure virial black hole
masses ($M_{\rm BH}$) in these objects may be problematic, as the
method relies on the assumption that the emitting gas is in virial
motion around the black hole.

To investigate the structure in this region, we turn to reverberation
mapping (\citealt{Blandford82}, \citealt{Peterson93}). This method has
been extensively used to measure the physical size of the broad line
region (BLR) in Type 1 AGNs, including NLS1s. Reverberation mapping
relies on the correlation between variations of the AGN continuum
emission and the subsequent response of the broad emission lines that
are seen in Type 1 AGNs. By monitoring AGN spectra over a period of
time, one can measure the radius of the emitting gas from the central
source by observing a time delay between variations in the continuum
and emission line fluxes. Assuming the gas is in virial motion, this
radius can be combined with BLR gas velocity dispersion estimates to
obtain a measurement of $M_{\rm BH}$. To date, this method has been
applied to measure BLR radii in nearly 50 AGNs
(e.g., \citealt{Peterson04}, \citealt{Denney10}, \citealt{Bentz10}).

If broad high-ionization lines like \HeII \ are emitted from
virialized gas near the black hole, we expect much shorter
reverberation time lags for \HeII \ than for low ionization lines like
H$\beta\lambda4861$ because both ionization stratification and the
line width require this gas to be much closer to the central
source. \cite{Peterson00b} investigated the NLS1 NGC\,4051 and
detected very broad, blue-enhanced \HeII \ emission in the RMS
spectrum. Unfortunately, their time resolution was inadequate to
reliably measure a \heii \ lag. \cite{Denney09b} measured an improved
\Hbeta \ lag for NGC\,4051 using data from their 2007 campaign, but
were similarly unable to recover a \heii \ lag. \cite{Bentz10} report
marginal detections of \heii \ lags in two NLS1 galaxies observed in
their 2008 observing campaign. These recent studies achieved similar
high sampling rates, but the expected \heii \ lags for these targets
are too short for a robust detection in these datasets. We have
therefore been unable to test whether this emission originates in
outflowing gas or is in virial motion.

We included the NLS1 galaxy Mrk 335 in a recent reverberation mapping
campaign that will be described in detail elsewhere (Grier et al., in
preparation). One goal of this high sampling rate program was to
measure the reverberation lag for a high-ionization line, \HeII, in
this source. Here we present the \heii \ results, having obtained a
high enough sampling rate to measure its short time delay, and show
that the \mbh estimate from the high ionization \heii \ line agrees
with those from low ionization lines.

\section{OBSERVATIONS}
In general, we follow the observational and analysis practices of
\cite{Denney10}, which largely follows the analysis described by
\cite{Peterson04}. Details on observations and subsequent analysis
techniques will be discussed in the accompanying work by Grier et
al.

\subsection{Spectroscopy}
\begin{figure}
\begin{center}
\epsscale{1.0}
\plotone{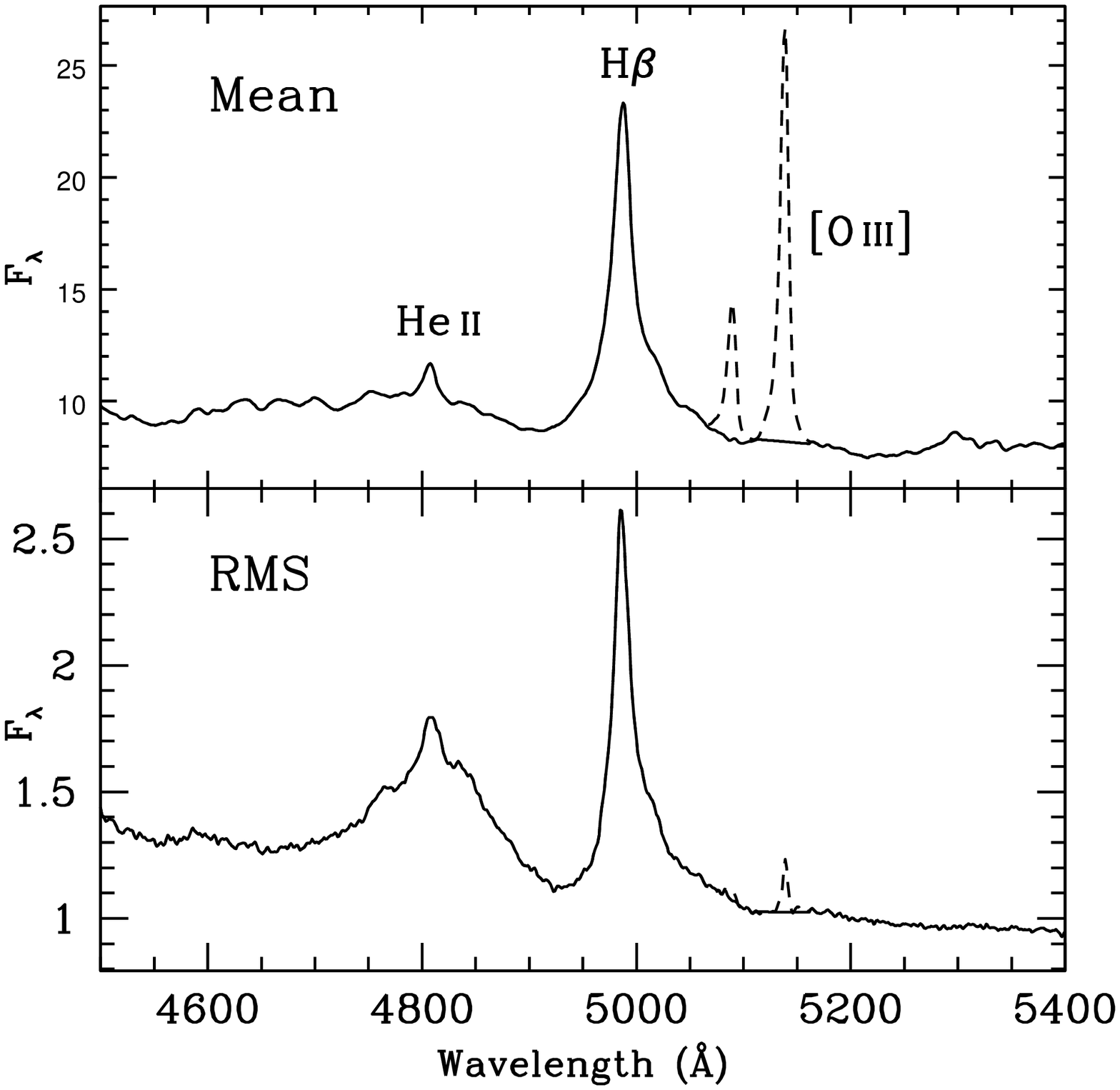}
\caption{The flux-calibrated mean (top) and RMS (bottom) spectra of
Mrk335 in the observed frame ($z$ = 0.02579). The flux density is in
units of 10$^{-15}$ erg s$^{-1}$ cm$^{-2}$ \AA$^{-1}$. The dashed
lines show the spectra before the [\oiii] \ narrow emission lines were
removed; the solid line shows the spectra after the subtraction.}
\label{fig:f1}
\end{center}
\end{figure}

The majority of the spectra were obtained using the 1.3m McGraw-Hill
telescope at MDM Observatory. We used the Boller and Chivens CCD
spectrograph to obtain 82 spectra over the course of 120 nights from
2010 Aug 31 to Dec 28. We used the 350 mm$^{-1}$ grating to obtain a
dispersion of 1.33 \AA\,pixel$^{-1}$, with a central wavelength of
5150\,\AA \ and overall spectral coverage from roughly 4400\,\AA \ to
5850\,\AA. The slit was oriented north-south and set to a width of
$5''\!.0$ and we used an extraction window of $12''\!.0$, which
resulted in a spectral resolution of 7.9 \AA. Figure \ref{fig:f1}
shows the mean and root mean square (RMS) spectra of Mrk 335 from the
MDM spectra.

We also obtained 7 spectra with the Nasmith spectrograph and SPEC-10
CCD at the 2.6m Shajn telescope at the Crimean Astrophysical
Observatory (CrAO). We used a $3''\!.0$ slit at a position angle of
90$^{\circ}$. Spectral coverage was from approximately 3900 \AA \ to
6100 \AA.

\subsection{Photometry}
We collected 25 epochs of $V$-band photometry from the 70-cm telescope
at CrAO using the AP7p CCD at prime focus, covering a 15\arcmin
$\times$15\arcmin \ field of view. The flux was measured within an
aperture of $15''\!.0$. See \cite{Sergeev05} for more details.

We also obtained 19 epochs of $V$-band photometry at the Wise
Observatory of Tel-Aviv University using the Centurion 18-inch
telescope with a 3072 $\times$ 2048 STL-6303E CCD and a field of view
of 75\arcmin$\times$50\arcmin. For these data, we used the ISIS image
subtraction package rather than aperture photometry to measure fluxes
(\citealt{Alard98}; \citealt{Alard00}) following the procedure of
\cite{Shappee11}.

\section{LIGHT CURVES AND TIME SERIES ANALYSIS}
\subsection{Light Curves}
The reduced spectra were flux-calibrated assuming that the [O\,{\sc
iii}]\,$\lambda5007$ emission line flux is constant (see
\citealt{Denney10} for details on data processing). Emission-line
light curves were created for both the MDM and CrAO data sets by 
fitting linear continua underneath the \Hbeta \ and \heii \ lines and 
integrating the flux above them. \Hbeta \ fluxes were measured between
4910--5100\,\AA \ in the observed frame, with the continuum
interpolation defined by the regions 4895--4910 and 5215--5240
\AA. \HeII \ fluxes were measured from 4660--4895\,\AA, with the
continuum defined from 4550--4575 and 4895--4910\,\AA. The CrAO light
curves were then scaled to the MDM light curve to account for the
different amounts of [O\,{\sc iii}] light that enters the slits due to
differences in seeing, slit orientation, and aperture size.

\begin{figure}
\begin{center}
\epsscale{1.0} 
\plotone{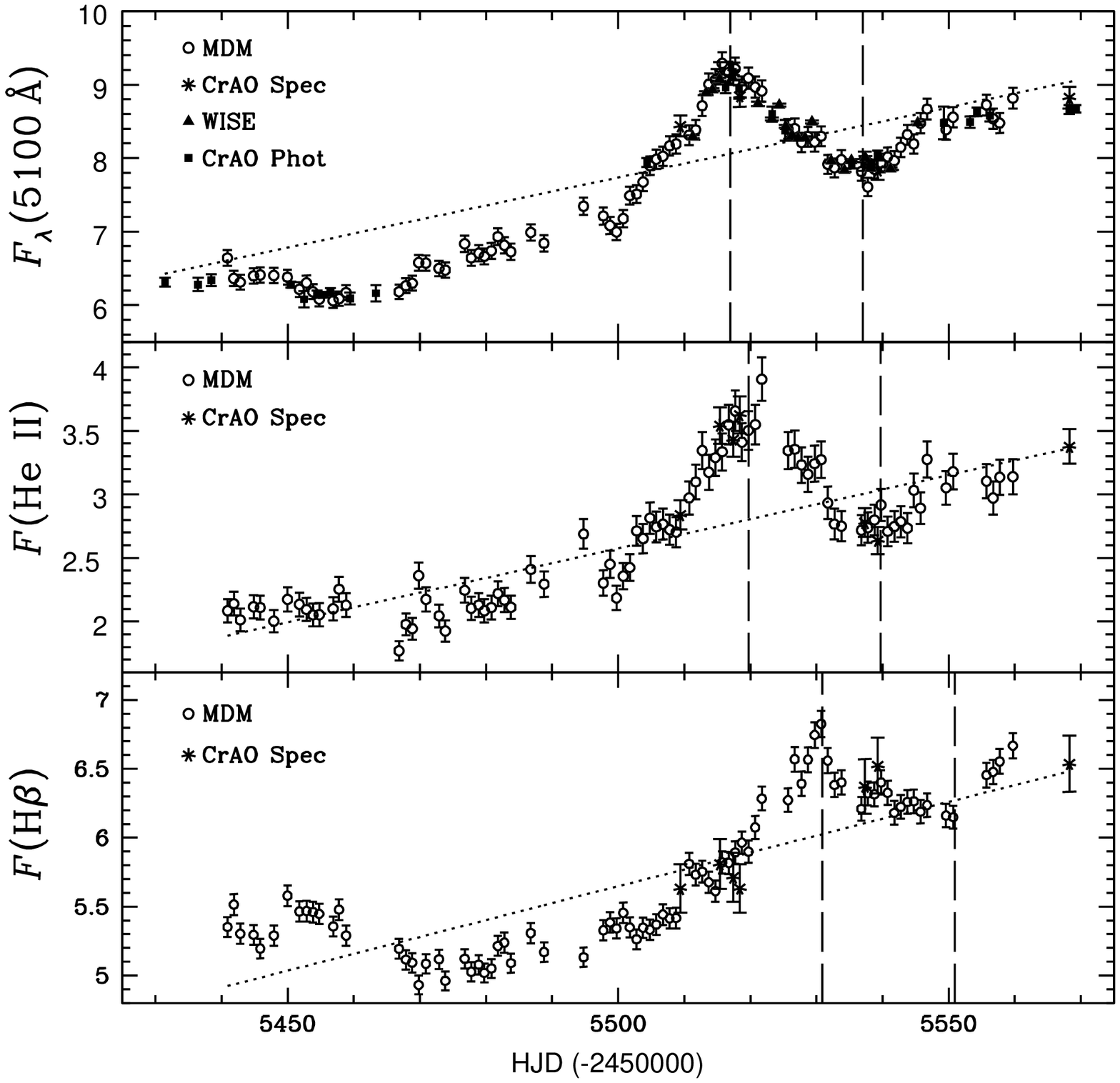}
\caption{Complete light curves for Mrk 335 from our observing
campaign. The top panel shows the 5100\AA \ flux in units of
$10^{-15}$ \ergscm\AA$^{-1}$, the middle panel shows the flux in the
\HeII \ region in units of $10^{-13}$ \ergscm, and the bottom panel
shows the integrated H$\beta\lambda4861$ flux, also in units of
$10^{-13}$ \ergscm. Open circles denote observations from MDM
Observatory and asterisks represent spectra taken at CrAO. Closed
squares show the photometric observations from CrAO, and closed
triangles represent photometric observations from WISE
Observatory. Vertical dashed lines have been placed at two obvious
features in the continuum to aid the eye. The vertical lines have been 
shifted by the measured \heii \ and \Hbeta \ lag values (2.7 days and
13.9 days, respectively) to aid the eye in identifying the correct lag
values for each emission line. Dotted lines show the trends that were
subtracted before performing the cross correlation analysis.}
\label{fig:f2}
\end{center}
\end{figure}

The continuum light curve was created by taking the average 5100\,\AA
\ continuum flux of the MDM spectra, measured between 5215-5240\,\AA \
in the observed frame. This light curve was then scaled and merged
with the other continuum and photometric light curves with corrections
for the host galaxy starlight in the different apertures (see
\citealt{Peterson95}). The final continuum and emission-line light
curves are shown in Figure \ref{fig:f2}. Light curve statistics are
given in Table \ref{Table:tbl1}.

\subsection {Time delay measurements}

\begin{figure}
\begin{center}
\epsscale{1.0} 
\plotone{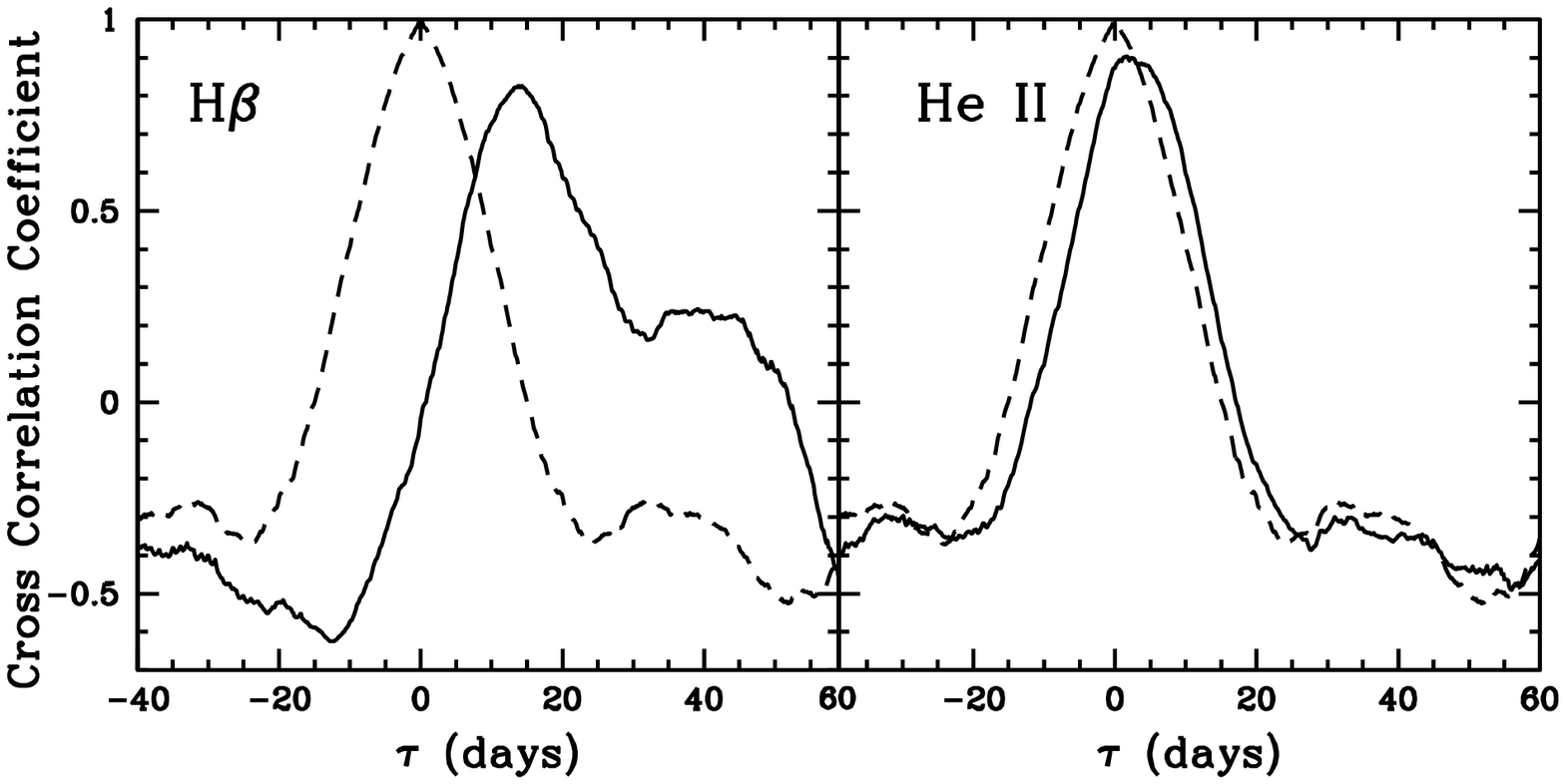}
\caption{CCFs for the light curves. Dashed lines represent the
autocorrelation function (ACF) of the continuum light curve, and solid
lines show the CCFs for the emission lines.}
\label{fig:f3}
\end{center}
\end{figure}

For comparison with previous results, we used the interpolation method
originally described by \cite{Gaskell86} and \cite{Gaskell87} which
was later modified by \cite{White94} and Peterson et al.\,(1998, 2004)
to measure the time lag. We cross-correlated the two light curves with
one another, calculating the value of the cross correlation
coefficient $r$ at each value of time lag. Figure \ref{fig:f3} shows
the resulting cross correlation functions (CCFs) for the light
curves. Uncertainties in lags are calculated using Monte Carlo
simulations that employ the methods of \cite{Peterson98} and refined
by \cite{Peterson04}. For each realization, we measure the location of
the peak value of the cross correlation coefficient ($\tau_{\rm
peak,CCF}$), and the centroid of the CCF ($\tau_{\rm cent,CCF}$),
calculated using points surrounding the peak. We adopt the mean
$\tau_{\rm peak,CCF}$ and $\tau_{\rm cent,CCF}$ from the Monte Carlo
realizations for our delay measurements and the standard deviation as
our formal uncertainties. Before the light curves were cross
correlated, we removed the long-term linear upward trend that is
clearly visible in all three light curves (see Figure
\ref{fig:f2}). \cite{Welsh99} discusses the value in this practice of
``detrending'' the light curves, as the cross correlation function
(CCF) tends to latch onto long-term trends unassociated with
reverberation, often resulting in incorrect lags. From our cross
correlation analysis, we measure $\tau_{\rm cent,CCF}$(\Hbeta)~=~13.9
$\pm$ 0.9 days and $\tau_{\rm cent,CCF}$(\heii)~=~2.7 $\pm$ 0.6
days. All lag measurements are listed in Table \ref{Table:tbl2}. 

Previous reverberation studies have relied on these fairly simple
cross correlation methods to measure $\tau$. Recently, however,
\cite{Zu11} discussed an alternative method of measuring reverberation
time lags called the Stochastic Process Estimation for AGN
Reverberation (SPEAR) and demonstrated its ability to recover accurate
time lags. The basic idea is to assume all emission-line light curves
are scaled and shifted versions of the continuum light curve. One then
fits the light curves using a damped random walk model
(e.g. \citealt{Kelly09}, \citealt{Kozlowski10}, \citealt{MacLeod10})
and then aligns them to determine the time lag. Uncertainties in lags
are computed using a Markov Chain Monte Carlo method (see
\citealt{Zu11}). SPEAR is remarkably good at predicting time lags in
data sets with relatively large gaps in the sampling. Using SPEAR, we
successfully recover time lags for both the \Hbeta\ and \heii\
emission lines. We allowed SPEAR to automatically remove the linear
trend and include any resulting uncertainties in the overall lag
uncertainties. We measure $\tau_{\rm SPEAR}$(\Hbeta)~=~14.0~$\pm$~0.3
days and $\tau_{\rm SPEAR}$(\heii)~=~$1.6^{+0.7}_{-0.5}$ days, also
reported in Table \ref{Table:tbl2}. We see good agreement with the CCF
results from the \Hbeta \ emission line, and while there is a small
difference between SPEAR and CCF lags for the \heii \ line, they are
still statistically consistent with one another. We suspect this small
difference is due to the gap in data that is very close to the peak in
the \heii \ light curve. We adopt the CCF values for our mass
calcuations to allow comparison with previous reverberation efforts.

\subsection {Line width measurement and \mbh calculations}
Assuming that the motion of the H$\beta$-emitting gas is dominated by
gravity, the relation between $M_{\rm BH}$, line width, and time delay
is
\begin{equation}
M_{\rm BH} = \frac{f c \tau \Delta V^2}{G},
\end{equation}
where $\tau$ is the measured emission-line time delay, $\Delta V$ is
the velocity dispersion of the BLR, and $f$ is a dimensionless factor
that accounts for the structure within the BLR. The BLR velocity
dispersion can be estimated using the line width of the measured broad
emission line in question. This width is usually characterized by
either the FWHM or the line dispersion, $\sigma_{\rm line}$. We use
$\sigma_{\rm line}$ because there is evidence that it produces less
biased \mbh measurements (\citealt{Peterson11}). We measure the line
width in the RMS spectrum, which eliminates contributions from the
contaminating narrow components. We adopt an average value of
$<f>=5.5$ based on the assumption that AGNs follow the same
\mbh--\sigstar relationship as quiescent galaxies
(\citealt{Onken04}). This is consistent with \cite{Woo10} and allows
easy comparison with previous results, but is about a factor of two
larger than the value of $<f>$ computed by \cite{Graham11}.

To determine the best value of $\sigma_{\rm line}$, we use Monte Carlo
simulations following \cite{Peterson04}. The resulting line widths are
given in Table \ref{Table:tbl2}. Using our measured values of
$\tau_{\rm cent,CCF}$ for the average time lag and $\sigma_{\rm line}$
from the RMS spectrum as $\Delta V$, we compute \mbh using both the
\Hbeta \ and \heii \ emission lines. We measure \mbh = ($2.7 \pm 0.3
$)$ \times 10^{7} \ M_{\odot}$ using the \Hbeta \ emission line and
\mbh = ($2.6 \pm 0.6 $)$ \times 10^{7} \ M_{\odot}$ using \heii.
 
\section{DISCUSSION}
As discussed above, several studies involving NLS1 galaxies have found
indications of outflows in high-ionization lines in the form of
enhanced flux on the blue side of the emission lines. Inspection of
Mrk 335 spectra from the $HST$ archive shows this enhanced blueward
flux is present in the \civ$\lambda$1549 line as well, but the
\heii$\lambda$1640 line is blended with \civ, so we cannot see if it
too exhibits this blue enhancement. In fact, the shape of the \HeII \
emission line in the RMS spectrum of Mrk 335 (Figure \ref{fig:f1}) shows red
and blue shoulders that could be a signature of disk structure. To
search for possible outflow signatures in the \HeII \ emission line,
we divided it into red and blue components and integrated each
component separately, creating two \heii \ light curves. We then
cross-correlated the red and blue light curves with one another to see
if there is any time delay between the two components. Cross
correlation analysis yields a centroid lag $\tau_{\rm
cent}$~=~0.4~$\pm$~0.8 days. This is consistent with zero and thus
presents no evidence for bulk outflows in the \HeII \ emission of Mrk
335. The consistency of the \mbh measurements made using the \heii \
lines with those from \Hbeta \ are also suggestive of virial motion
rather than outflowing gas.

Previous reverberation measurements of Mrk 335 were made by
\cite{Kassebaum97} and \cite{Peterson98} and subsequently reanalyzed
by \cite{Peterson04} and \cite{Zu11}. \cite{Zu11} report a time delay
of 15.3$^{+3.6}_{-2.2}$ days and $\sigma_{\rm line}$ $\sim$ 920 km
s$^{-1}$ for \Hbeta, but were unable to make a robust \heii \
measurement, as their average time sampling was on the order of 10
days. \cite{Peterson04} measure \mbh = ($1.4 \pm 0.4 $)$ \times 10^{7}
\ M_{\odot}$ from the \Hbeta \ emission line. Our \mbh measurements
deviate from theirs by almost a factor of two. We suspect the
difference in \mbh is due to the difference in the line width
measurements between the two campaigns. The uncertainties in line
width measurements and in the $f$ factor are the main sources of
uncertainties in reverberation \mbh measurements -- when the light
curves are well-sampled, the lag measurements themselves have been
shown to be remarkably robust
(e.g. \citealt{Watson11}). \cite{Peterson04} (Table 6) find that the
virial products computed for an object using data from different
epochs often differ from one another by as much as a factor of two
(e.g. NGC\,5548) and that the typical fractional error in the virial
products is about 33\%. Given our uncertainties in the $f$ factor and
the limitations in trying to accurately describe the BLR velocity
field with a single line-width characterization, we probably cannot
actually do better than about a factor of two or three in individual
\mbh measurements.

Ground-based reverberation campaigns in the past have been limited to
objects with \Hbeta \ time lags that are expected to be less than a
month or two due to both the finite length of the campaigns (which
typically last 50--100 days) and the fact that most objects are only
observable from the ground for only about half of the
year. Measurement of longer time lags would require extended
campaigns, which are difficult to schedule. If, as our evidence
suggests, the \heii \ emission line is in virial motion around the
black hole, we can use this emission line to measure \mbh in objects
at higher redshifts, as expected \heii \ lags in many of these
high-luminosity objects are short enough to measure in one observing
season.

\section{SUMMARY}
We have presented the first robust \HeII \ reverberation lag
measurement in a NLS1 galaxy. We also measure the \Hbeta \ time lag in
this galaxy and compute \mbh using both emission lines. The \mbh
measurements from \heii \ and \Hbeta \ are consistent with one
another, suggesting that the gas producing the \heii \ emission
resides in the same structure as that producing \Hbeta\
emission. While other high-ionization lines such as \civ \ show
evidence for outflows, we do not see this in \heii, possibly because
the \heii-emitting gas does not arise cospatially with gas producing
\civ \ emission. This has practical implications for future
reverberation efforts, as the \heii \ emission may allow us to more
efficiently measure \mbh in objects at high redshift.

\acknowledgments 
We gratefully acknowledge the support of the National
Science Foundation through grant AST-1008882. BJS, CBH, and JLV are
supported by NSF Fellowships. CSK and DMS acknowledge the support of
NSF grant AST-1004756. AMM acknowledges the support of Generalitat
Valenciana, grant APOSTD/2010/030. SK is supported at the Technion by
the Kitzman Fellowship. SK and SR are supported by a grant from the
Israel-Niedersachsen collaboration program. SR is also supported at
Technion by the Lady Davis Fellowship. SGS acknowledges the support to
CrAO in the frame of the 'CosmoMicroPhysics' Target Scientific
Research Complex Programme of the National Academy of Sciences of
Ukraine (2007-2012). VTD acknowledges the support of the Russian
Foundation of Research (RFBR, project no. 09-02-01136a). The CrAO CCD
cameras were purchased through the US Civilian Research and
Development for Independent States of the Former Soviet Union (CRDF)
awards UP1-2116 and UP1-2549-CR-03. This research has made use of the
NASA/IPAC Extragalactic Database (NED) which is operated by the Jet
Propulsion Laboratory, California Institute of Technology, under
contract with the National Aeronautics and Space Administration.


\clearpage

\begin{deluxetable}{lccccccc}
\tablewidth{0pt}
\tablecaption{Light Curve Statistics}
\tablehead{
\colhead{ } &
\colhead{ } &
\multicolumn{2}{c}{Sampling} &
\colhead{ } &
\colhead{Mean} \\
\colhead{Time} &
\colhead{ } &
\multicolumn{2}{c}{Interval (days)} &
\colhead{Mean} &
\colhead{Fractional} \\
\colhead{Series } &
\colhead{$N$} &
\colhead{$\langle T \rangle$} &
\colhead{$T_{\rm median}$} &
\colhead{Flux} &
\colhead{Error} &
\colhead{$F_{\rm var}$} &
\colhead{$R_{\rm max}$} \\
\colhead{(1)} &
\colhead{(2)} &
\colhead{(3)} &
\colhead{(4)} &
\colhead{(5)} &
\colhead{(6)} &
\colhead{(7)} &
\colhead{(8)} 
} 
\startdata
5100\,\AA\              & 133 & 1.0 & 0.95  & $7.57\pm1.01$  & 0.013 & 0.129 & $1.53\pm 0.04$\\
\HeII                   & 89  & 1.5 & 1.00  & $7.65\pm1.00$  & 0.016 & 0.130 & $1.55\pm 0.04$\\
H$\beta$$\lambda$4861   & 89  & 1.5 & 1.00  & $5.74\pm0.53$  & 0.015 & 0.091 & $1.38\pm 0.03$
\enddata
\tablenotetext{*}{Column (1) lists
the spectral feature, and column (2) gives the number of points in the
individual light curves. Columns (3) and (4) list the average and
median time spacing between observations, respectively. Column (5)
gives the mean flux of the feature in the observed frame , and column
(6) shows the mean fractional error that is computed based on
observations that are closely spaced in time. Column (7) gives the
excess variance, defined by
\begin{equation}
F_{\rm var} = \frac{\sqrt{\sigma^{2}-\delta{^2}}}{\langle f\rangle}
\end{equation}
where $\sigma^{2}$ is the flux variance of the observations,
$\delta^{2}$ is the mean square uncertainty, and $\langle f \rangle$
is the mean observed flux. Column (8) is the ratio of the maximum to
minimum flux in each light curve.} 
\tablenotetext{*}{Continuum and emission-line fluxes 
are given in $10^{-15}$ \ergscm\AA$^{-1}$ and $10^{-13}$ \ergscm, 
respectively.}
\label{Table:tbl1}
\end{deluxetable} 

\begin{deluxetable}{lcccc} 
\tablewidth{0pt} 
\tablecaption{\Hbeta \ and \HeII \ Time Series Results} 
\tablehead{ 
\colhead{Parameter} & 
\colhead{\Hbeta} & 
\colhead{\heii} \\ 
\colhead{(1)} & 
\colhead{(2)} & 
\colhead{(3) }  
} 
\startdata
$\tau_{\rm cent,CCF}$\tablenotemark{a}	      &  13.9  $\pm$ 0.9 days        &  2.7 $\pm$ 0.6 days \\
$\tau_{\rm peak,CCF} $\tablenotemark{b}	      &  13.8  $\pm$ 0.8 days        &  2.1 $\pm$ 1.2 days  \\
$\tau_{\rm SPEAR}$            &  14.0  $\pm$ 0.3 days        &  1.6 $^{+0.7}_{-0.5}$ days  \\		   
$\sigma_{\rm line}$ (mean)    &  1641  $\pm$ 12 km s$^{-1}$  &  3465 $\pm$ 26 km s$^{-1}$ \\
FWHM (mean)                   &	 1363  $\pm$ 15 km s$^{-1}$  &  3191 $\pm$ 571 km s$^{-1}$ \\
$\sigma_{\rm line}$ (RMS)     &	 1336  $\pm$ 51 km s$^{-1}$  &  3001 $\pm$ 277 km s$^{-1}$ \\
FWHM (RMS)	              &	 1149  $\pm$ 38 km s$^{-1}$  &  7380 $\pm$ 1275 km s$^{-1}$ \\
$M_{\rm BH}$                  &  ($2.7 \pm 0.3 $)$ \times 10^{7} M_{\odot}$& ($2.6 \pm 0.8 $)$ \times 10^{7} M_{\odot}$
\enddata
\tablenotetext{*}{All values are given in the rest frame of the object.}
\label{Table:tbl2}
\end{deluxetable} 

\end{document}